# Comparative Study of Dense Bulk MgB$_2$ Materials Prepared by Different Methods

V.N. Narozhnyi[1,2], G. Fuchs[1], A. Handstein[1], A. Gümbel[1], J. Eckert[1], K. Nenkov[1], D. Hinz[1], O. Gutfleisch[1], A. Wälte[1], L.N. Bogacheva[2], I.E. Kostyleva[2], K.-H. Müller[1], L. Schultz[1]

**Received**

We report on the results of a comparative investigation of highly dense bulk MgB$_2$ samples prepared by three methods: (i) hot deformation; (ii) high pressure sintering; and (iii) mechanical alloying of Mg and B powders with subsequent hot compaction. All types of samples were studied by *ac*-susceptibility, *dc*-magnetization and resistivity measurements in magnetic fields up to $\mu_0 H = 160$ kOe. A small but distinct anisotropy of the upper critical field $H_{c2}^{a,b}/H_{c2}^{c} \sim 1.2$ connected with some texture of MgB$_2$ grains was found for the hot deformed samples. The samples prepared by high pressure sintering as well as by mechanical alloying show improved superconducting properties, including high upper critical fields $H_{c2}$ ($\mu_0 H_{c2}(0) \sim 23$ T), irreversibility fields $H_{irr}$ which are strongly shifted towards higher values $H_{irr}(T) \sim 0.8 H_{c2}(T)$ and high critical current $J_c$ ($J_c = 10^5$ A/cm$^2$ at 20 K and 1 T).

Soon after the discovery of superconductivity in MgB$_2$ it was recognized that the irreversibility fields $H_{irr}(T)$ of the samples prepared by standard solid state reaction are about two times smaller than the upper critical fields $H_{c2}(T)$, see e.g. [1,2]. All possible practical applications based on non-vanishing critical current densities $J_c$ are restricted to magnetic fields $H$ lower than $H_{irr}(T)$. It was also shown that for MgB$_2$, contrary to high-$T_c$ cuprates, $J_c$ is not limited by grain boundaries [3]. Considerable porosity of the samples prepared by standard technique is one of the factors limiting $J_c$. An increase of $H_{irr}(T)$ as well as an improvement of $J_c$ of MgB$_2$ can be achieved by preparation of highly dense material [4-7]. In this paper we report on a comparative investigation of highly dense bulk MgB$_2$ samples prepared by three different methods: (i) hot deformation; (ii) high

___

[1]Leibniz-Institut für Festkörper- und Werkstoffforschung Dresden, PO Box 270116, D-01171 Dresden, Germany

[2]Department of Low Temperature Physics, Institute for High Pressure Physics Russian Academy of Sciences, Troitsk, Moscow Reg., 142190, Russia

pressure sintering; and (iii) mechanical alloying of Mg and B powders with subsequent hot compaction.

Hot deformed samples were prepared by die-upsetting in an Ar atmosphere at temperatures from 850 to 900°C resulting in a height reduction of the initially prepared (by conventional solid-state reaction from B powder and Mg) MgB$_2$ pellets of about 70% [5]. High-pressure high-temperature sintering was performed using commercial Alfa Aesar MgB$_2$ powder at temperatures 800–1000°C and pressures 3–8 GPa using a "Toroid" high pressure cell [6]. Nanocrystalline samples were obtained by milling of Mg and B powders in Ar-atmosphere using a tungsten carbide (WC) vial containing WC balls with subsequent hot compaction at 700°C and 640 MPa [7]. All types of the samples were studied by *ac*-susceptibility, *dc*-magnetization and resistivity measurements in magnetic fields up to $\mu_0 H = 160$ kOe.

X-ray diffraction studies revealed a weak alignment of the c-axis of hexagonal MgB$_2$ grains along the force direction for the hot deformed samples. At the same time no texture was found for high pressure sintered material. The comparison of the diffraction patterns of the samples sintered under high pressure and the starting $MgB_2$ powder showed the appearance of additional weak reflections connected with MgO impurity in the sintered samples. For the ball-milled samples, beside a small amount of MgO there is visible only a little contamination of about 0.3 vol. % WC stemming from the milling tools. The broad diffraction peaks indicate small dimensions of coherent scattering domains of about 15 nm [7]. The microstructures of the investigated samples are considerably different. For the hot deformed samples, some orientation of the relatively flat $MgB_2$ grains was observed by optical investigation. This orientation corresponds to the texture revealed by x-ray diffraction. No preferable orientation has been found for high pressure sintered material. The average size of grains is about 20-50 μm for hot deformed samples as well as for samples sintered at high pressure. The mechanically alloyed and hot compacted nanocrystalline samples have about 1000 times smaller grains. Scanning electron microscope images mainly show spherical grains of about 40-100 nm in size, larger than the domain size evaluated from x-ray analysis. This indicates a significant amount of strain and structural defects [7]. The densities of all of these samples are close to the theoretical one.

A small, but distinct anisotropy of the upper critical fields $\gamma = H_{c2}^{a,b}/H_{c2}^{c} \sim 1.2$ was found for hot deformed samples by resistance and susceptibility measurements, see Fig. 1. Obviously this anisotropy is connected with the texture discussed above. Although the observed anisotropy of $H_{c2}(T)$ is considerably smaller than reported for $MgB_2$ single crystals ($\gamma \approx$ 1.7 - 6, see e.g. [8]) it probably can be increased by improvement of the preparation technique.

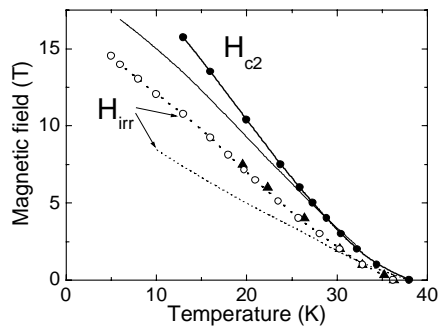

**Fig. 2.** $H_{c2}(T)$ and $H_{irr}(T)$ of $MgB_2$ sintered under high pressure. $H_{irr}$ data derived from the imaginary part of the *ac*-susceptibility are marked by triangles and are comparable with the resistively determined data marked by open circles. $H_{c2}$ and $H_{irr}$ of the bulk sample prepared by standard method [3] are shown as full and dotted lines, respectively, without symbols.

Fig. 2 shows the upper critical field and the irreversibility line of the sample sintered at high pressure. $H_{c2}(T)$ and $H_{irr}(T)$ were determined at 90 % of the normal-state resistance and zero resistance, respectively. These fields agree well with the values of $H_{c2}(T)$ determined from the onset of the superconducting transition from *dc*-magnetization measurements as well as with $H_{irr}(T)$ received from (1) the points separating reversible and irreversible regions of the *H-T* phase diagram determined by *dc*-magnetization and (2) from the peak value of the imaginary part of the *ac*-susceptibility [3,6]. The optimum conditions for high-pressure sintering were determined to be 1000°C and 3 GPa [6]. The values of $H_{c2}(T)$ and $H_{irr}(T)$ for samples sintered under high pressure as well as for nanocrystalline samples (Fig. 3) are considerably higher than those for the samples prepared by standard technique; $H_{c2}(0)$ can be estimated as 23 T and 22 T, respectively (18 T for standard method). Even more important is the pronounced shift of the irreversibility lines from $H_{irr}(T) \sim 0.5 H_{c2}(T)$ for standard bulk samples to $H_{irr}(T) \sim 0.7 - 0.8 H_{c2}(T)$ for high pressure sintered as well as for nanocrystalline samples, see Figs. 2,3.

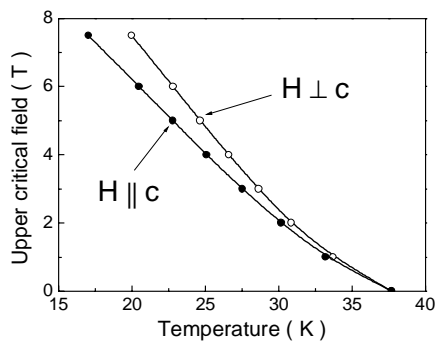

**Fig. 1.** Temperature dependence of the upper critical field of a textured $MgB_2$ sample for two directions of *H*.

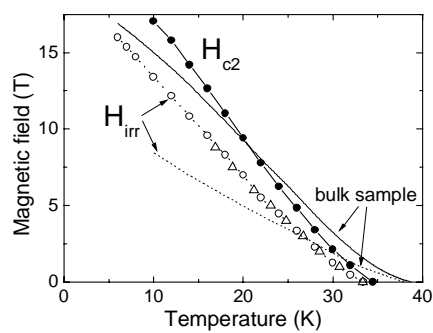

**Fig. 3.** Temperature dependence of $H_{c2}$ and $H_{irr}$ of nanocrystalline $MgB_2$ sample. For further details see the caption of Fig. 2.

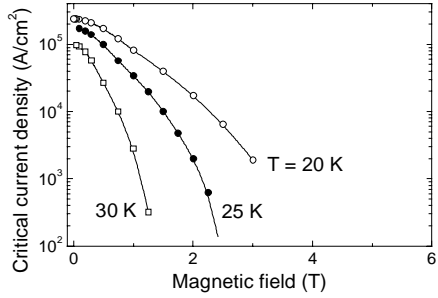

**Fig. 4.** Field dependence of the critical current density of textured $MgB_2$ for magnetic fields $H \parallel c$.

$J_c(H)$ curves determined from magnetization loops for all three types of samples are shown in Figs. 4-6. Although at $T = 20$ K the largest $J_c(H)$ ($\geq 10^5$ A/cm$^2$ at $H \leq 2$ T) was obtained for nanocrystalline material, at higher temperatures ($T = 30$ K) $J_c(H)$ is considerably higher for high pressure sintered sample. This is connected with the reduced $T_c = 34.5$ K of the nanocrystalline samples in comparison with $T_c = 38$ K for samples prepared by high-pressure sintering.

Improved flux pinning manifesting itself in the observed increased $H_{irr}(T)$ as well as in the high $J_c(H)$ can be connected with small grains and the enhanced number of grain boundaries for the nanocrystalline material. At the same time comparable characteristics obtained on samples prepared under high pressure indicate that some defects inside the grains should play the role of effective pinning centers. Further increasing of $MgB_2$ parameters can be expected by improvement of the preparation technique.

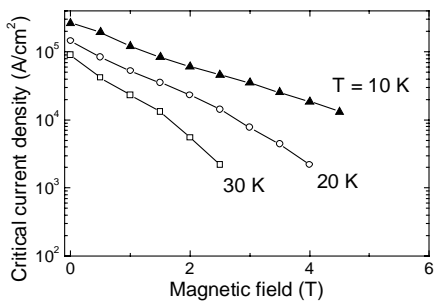

**Fig. 5.** Field dependence of the critical current density of $MgB_2$ sintered under high pressure for $T = 10$, 20 and 30 K.

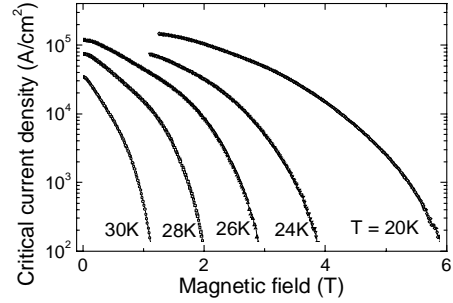

**Fig. 6.** Field dependence of the critical current density of nanocrystalline $MgB_2$ at $T = 20, 24, 26, 28$ and 30 K.

## ACKNOWLEDGMENTS

This work was partially supported by DFG (grant MU1015/8-1) and RFBR (grant 01-02-04002).

## REFERENCES

1. D.K. Finnemore, J.E. Ostenson, S.L. Bud'ko, G. Lapertot, and P.C. Canfield, *Phys. Rev. Lett.* **86**, 2420 (2001).
2. G. Fuchs, K.-H. Müller, A. Handstein, K. Nenkov, V.N. Narozhnyi, D. Eckert, M. Wolf, and L. Schultz, *Sol. St. Commun.* **118**, 497 (2001).
3. D.C Larbalestier, L.D. Cooley, M.O. Rikel, A.A. Polyanskii, J. Jiang, S. Patnaik, X.Y. Cai, D.M. Feldman, A. Gurevich, A.A. Squitieri, M.T. Naus, C.B. Eom, E.E. Hellstrom, R.J. Cava, K.A. Regan, N. Rogadao, M.A. Hayward, T. He, J.S. Slusky, P. Khalifah, K. Inumaru, and M. Haas, *Nature* **410**, 186 (2001).
4. Y. Takano, H. Takeya, H. Fujii, H. Kumakura, T. Hatano, and K. Togano, *Appl. Phys. Lett.* **78**, 2914 (2001)
5. A. Handstein, D. Hinz, G. Fuchs, K.-H. Müller, K. Nenkov, O. Gutfleisch, V.N. Narozhnyi, and L. Schultz, *J. Alloys Comp.* **329**, 285 (2001).
6. V.N. Narozhnyi, A. Handstein, K. Nenkov, G. Fuchs, L.N. Bogacheva, I.E. Kostyleva, and K.-H. Müller, Int. J. Mod. Phys. B (to be publ.).
7. A. Gümbel, J. Eckert, G. Fuchs, K. Nenkov, K.-H. Müller, and L. Schultz, *Appl. Phys. Lett.* **80**, 2725 (2002).
8. M. Xu, H. Kitazawa, Y. Takano, J. Ye, K. Nishida, H. Abe, A. Matsushita, and G. Kido, *Appl. Phys. Lett.* **79**, 2779 (2001).